# Hydrogenated anatase: Strong photocatalytic H$_2$ evolution without the use of a co-catalyst

Ning Liu[1], Christopher Schneider[1], Detlef Freitag[2], Umamaheswari Venkatesan[3], V. R. Reddy Marthala[3], Martin Hartmann[3], Benjamin Winter[4], Erdmann Spiecker[4], Eva Zolnhofer[5], Karsten Meyer[5], Patrik Schmuki[1]*

[1]Department of Materials Science WW-4, LKO, University of Erlangen-Nuremberg, Martensstrasse 7, 91058 Erlangen, Germany;
[2]High Pressure Laboratory, Chair of Separation Science and Technology, University of Erlangen-Nuremberg, Haberstrasse 11, 91058 Erlangen, Germany;
[3]ECRC - Erlangen Catalysis Resource Center, University of Erlangen-Nuremberg, Egerlandstrasse 3, 91058 Erlangen, Germany;
[4]Center for Nanoanalysis and Electron Microscopy (CENCM), University of Erlangen-Nuremberg, Cauerstrasse 6, 91058 Erlangen, Germany;
[5] Inorganic & General Chemistry, University of Erlangen-Nuremberg, Egerlandstrasse 1, 91058 Erlangen, Germany.
*Corresponding author. Tel.: +49 91318517575, fax: +49 9131 852 7582
Email: schmuki@ww.uni-erlangen.de

**Abstract:**

In the present work we show, how a high pressure hydrogenation of commercial anatase or anatase/rutile powder can create a photocatalyst for hydrogen evolution that is highly effective and stable without the need of any additional co-catalyst. This activation effect can not be observed for rutile. For anatase/rutile mixtures, however, a strong synergistic effect is found (similar to findings commonly observed for noble metal decorated $TiO_2$). ESR measurements indicate the intrinsic co-catalytic activation of anatase $TiO_2$ to be due to specific defect centers formed during hydrogenation.

Ever since the groundbreaking work of Fujishima and Honda in 1972 [1], $TiO_2$ is considered as a promising photocatalyst for the splitting of water into $H_2$ and $O_2$. In the original experiment, Fujishima et al. used a $TiO_2$ photoanode, connected via an external circuit to a platinum counter electrode – the latter was needed to successfully evolve $H_2$ from water. Due to the simplicity of the concept, illumination of a cheap and abundant semiconductor to create photoexcited charge carriers that can be transferred directly to water to form a high density energy fuel ($H_2$), the report found a tremendous scientific resonance. Meanwhile, more than 10000 papers have been published on using $TiO_2$ in a large palette of morphologies and modifications, to trigger a wide range of photocatalytic ractions (for overviews see e.g. refs. [2-8]). While numerous photoelectrochemical studies (i.e., using an illuminated $TiO_2$ electrode in an electrochemical circuit) where performed, still the most direct and economic approach is the use of $TiO_2$ in the form of particle suspensions – thus using the photocatalytic system without an external applied voltage. However, under these so called open-circuit conditions (OCP), $TiO_2$ alone is not efficient for the photoproduction of hydrogen without the use of a co-catalyst – mostly this is a noble metal (M), such as Pt, Pd or Au – for overviews see e.g. refs. [9-11]. These combined photocatalytic $M@TiO_2$ systems have therefore been widely investigated in view of optimizing their efficiency towards $H_2$ generation from water (with or without using sacrificial agents such as ethanol) [9, 12].

In general, the function of the noble metal co-catalyst has been described in terms of i) providing an electron acceptor that mediates electron transfer to the electrolyte, ii) forming solid state junctions (metal/semiconductor), or iii) acting as a hydrogen recombination center that strongly promotes $H_2$ formation [9-11]. In $M@TiO_2$ catalysts it has been generally observed that the crystalline phase of $TiO_2$ is a very important factor for the performance of such photocatalytic $H_2$-generation systems [6-9, 13, 14]. Anatase and rutile are the most commonly used polymorphs in photoactivated $TiO_2$ applications. In photocatalytic water splitting, generally M@anatase combinations are found to be more efficient than M@rutile.

This difference in photocatalytic activity is commonly attributed to a higher charge recombination rate of photoexcited carriers in rutile compared with anatase [9]. Nevertheless, compared to the pure phases of anatase or rutile, the presence of a mixed phase in M@TiO$_2$ catalysts was commonly found to lead to a considerable enhancement of the reaction rate. This synergistic effect has been attributed to grain internal junction formation (band-gap off-set of the two phases) [15], internal grain boundaries in mixed particles, or the interplay of the different phases with the noble metal particle decoration [2, 9].

For TiO2 based photocatalysts, large efforts are not only dedicated to accelerate reaction rates but also to extend the light absorption of TiO2 (Eg anatase = 3.2 eV; Eg rutile = 3.0 eV[13]) into the visible range of the spectrum to allow for a more efficient use of solar light.[2] Over the years, a large variety of band-gap engineering approaches were introduced, involving doping with a wide range of elements (for an overview, see e.g. refs.[2,8]). While the introduction of a number of extrinsic dopant states in the TiO2 band-gap creates a visible light activation, the effect on water splitting efficiencies remained for a long time modest. However, in 2011, Chen and Mao reported 'black' TiO2 particles that, when decorated with Pt, reached a very high open circuit water splitting activity − this was ascribed to a broad visible light absorption of 'black' TiO2.[12] This finding triggered a considerable amount of follow-up work and seemingly identical 'black' TiO2 has meanwhile been produced by a full range of conventional TiO2-reduction treatments (vacuum annealing, electrochemical reduction, Ar/H2 annealing under atmospheric conditions),[14-18] as well as the pure H2 treatment introduced by Chen and Mao[12]. Many of these reduction treatments of TiO2 are reported to lead to a black or dark-blue appearance of the material. The appearance of color in these established treatments is typically ascribed to oxygen vacancies or Ti3+ reductive defect states.[6, 18, 19] In contrast, Chen and Mao ascribed their finding to the formation of an amorphous layer at the outermost part of their TiO2 nanoparticles.[12, 20-22] In literature, thus, overall, for various reductive approaches, visible light absorption in TiO2 nanoparticles and a considerable number of effects is observed, independent of the exact nature of the treatment.

In the present work we show, however, a unique co-catalytic effect that is observed specifically for high-pressure H2 treated TiO2 anatase nanoparticles. Such a treatment can activate a strong and stable photocatalytic H2 evolution ability in commercial anatase or anatase/rutile nanoparticles, without the use of noble-metal co-catalysts. This activation is not observed if pure rutile powders are hydrogenated. Moreover, if conventional reduction processes, e.g. annealing in Ar or Ar/H2, are used for treating the polymorphs of TiO2, no significant activation can be found.

For our experiments we used a range of commercial anatase, rutile and mixed anatase/rutile powders, as specified in the experimental section and in the SI (Table S1 and Fig. S1). Typical hydrogenation was carried out in pure H2 at 500 °C and 20 bar pressure for durations between 1 h and 3 d. The photocatalytic H2-evolution activity of the different particles was evaluated by gas chromatography using suspensions of the particles in a H2O/methanol solution under AM 1.5 (100 mW/cm2) illumination (experimental details are given in the SI).

Fig. 1 shows the observed hydrogen evolution rate of the investigated anatase, rutile and mixed-phase samples with and without the hydrogenation treatment. After hydrogenation, clearly all anatase or anatase-containing samples show a significant photocatalytic H2 evolution, while for pure rutile samples no H2 generation could be detected. All untreated reference samples, anatase or rutile, did not show any detectable H2 evolution even for experiments carried out for up to 48 h of illumination. Characterization of the particles in view of their structure was carried out using X-ray diffraction (XRD) (Fig. 2a), and on selected samples using high-resolution transmission electron microscopy (HRTEM) and selected area electron diffraction (SAED) (SI, Fig. S2). The results show that all particles are well crystalline before and after the H2 treatment. The pure anatase samples do not show any conversion to rutile during the elevated temperature (500 °C) hydrogenation. This is in line with reports that normally phase transition from anatase to rutile TiO2 for nanocrystals requires annealing temperatures >600 °C.[23] For mixed phase particles we found, however, that the hydrogen treatment can cause the anatase content to decrease from 12 to approx. 4% (from XRD of sample M1 in SI Fig. S3). Both investigated mixed phase anatase-rutile samples exhibit higher photocatalytic activity than the pure anatase samples. For sample M2, the H2 evolution activity is more than twice the amount of the pure anatase sample. Regarding the influence of grain size on H2 evolution activity we compared An1 and An2 (particles An1 have a diameter of ≈25 nm while particles An2 have a diameter of ≈200 nm). Both lead, after hydrogenation, to a comparable H2 production rate; i.e., in our experiments we did not find the grain size, within the used variations, to be of a major factor. This may be ascribed to a commonly observed aggregation of TiO2 nanoparticles in neutral aqueous suspensions. Additional experiments were carried out to explore the influence of the hydrogenation duration. The results show that generally already after 1 h at 500 °C a significant activity for anatase is obtained, but longer time exposure of mixed oxide can increase the H2 evolution efficiency further (Fig. S4).

Moreover, this activation shows a remarkable stability. Fig. 1b shows hydrogen evolution as a function of time over a testing period of 5 days (measured in intervals of 8 h) using hydrogenated anatase (Hy-an1) as photocatalyst. Over the entire investigated time H2 is produced at a steady rate – no indication of activity-fading or poisoning can be observed.

In the literature, high-temperature hydrogen treatments of TiO2 are frequently described to reduce TiO2 in a similar manner as using other reductive gas environments, such as Ar/H2 or high temperature exposure to an inert gas. Therefore, except for the treatment in high pressure hydrogen, we tested Ar and Ar/H2 treatments using anatase powder (An1 and An2). The results in Fig. 1 show that for neither conventional reduction treatment a significant activation for H2 evolution could be observed.

Reference experiments using typical noble metal treatments on our anatase powder (Au-An1 and Pt-An1) show that at present the best measured beneficial co-catalytic effect of the H-treatment (Hy-an1) already achieves an efficiency of about 30% of a common Au nanoparticle co-catalyst or 10% of a common Pt treatment (see SI, Fig. S5). This is remarkable, considering that still a wide parameter range (H2 pressure, temperature, decoration, co-doping etc.) exists, that can be exploited for further optimization of the hydrogenated anatase treatments.

In order to explore changes induced by hydrogenation and elucidate the origin of the effect, we performed further XRD, TEM, X-ray photoelectron spectroscopy (XPS), solid-state proton nuclear magnetic resonance (1H MAS NMR), electron paramagnetic resonance (EPR) and photoluminescence (PL) investigations for anatase particles (An1) before and after the hydrogen treatment. In order to quantify the changes from XRD of An1 before and after the treatment, we carried out Rietveld refinement as given in Fig. 2b and SI S6. The results for the treated and untreated samples show a slight lattice parameter variation and clear shrinkage of the average nominal crystallite size from 32 to 22 nm. Such a reduction of crystallite size could be due to amorphization of the original lattice induced by the H2, if for instance as observed in literature [12,20] the amorphous shell around the particles would increase. However, our HRTEM of the hydrogenated particles did not show a significant increase of the thickness of these amorphous shells (Fig. 2c). Nevertheless, a change induced by hydrogenation becomes apparent for TEM taken under defocus (Fresnel contrast) conditions (Fig. 2d). As evident from the characteristic Fresnel contrast (bright dots in underfocus, dark dots in overfocus, see arrows), voids are present in the inside of the particles which may explain the reduced coherent volume (seemingly reduced diameter) that is obtained from the XRD (Rietveld refinement) data. Such voids are frequently observed in anatase particles after hydrogenation but rarely appear also in particles before treatment (see Fig. S7 in SI). Thus, hydrogenation seems to support the formation (or growth) of voids even though it is not the only origin of void formation. The appearance of voids inside of crystalline material may be due to internal gas bubble formation as known from metals[24], or more likely may be due to vacancy condensation [25,26]. In line with the XRD of Fig. 2a, the HRTEM images (SI, Fig. S2) of the void free parts of the crystal structure do not show a significant change in lattice parameters. In order to exclude beam induced effects under HRTEM conditions, we carried out various confirmation measurements (crystallization or void formation) – see e.g. SI, Fig. S8.

In order to further evaluate the influence of hydrogenation, solid-state 1H MAS NMR (SI, Fig. S9) and XPS (SI, Fig. S10) were carried out – both methods do not yield results that show a significant difference between the catalytically active and the non-active material. Nevertheless, it should be pointed out that due to the powder nature of the samples, 1H MAS NMR show strong broad peaks due to adsorbed water and -OH termination [27,28], that to a large extent originate from ambient uptake during sample transfer to the spectrometer. In spite of the heating experiment performed as described in the SI (that was carried out to remove adsorbed water, however a broad peak remains), i.e. it cannot be ruled out that for instance a signal from interstitial lattice hydrogen is present underneath the large signal originating from adsorbed water.

However, clear changes induced by hydrogenation can be well observed in EPR spectra and PL measurements. EPR spectra taken at room temperature and at 90 K under illumination are shown in Fig. 3a. The room temperature EPR spectrum of the non-hydrogenated anatase sample only shows a signature typically assigned to oxygen vacancies, i.e. does not indicate the presence of distinct Ti(III) related signals. At 90 K, spectra taken before and after UV illumination give a clear signature with orthorhombic g values of [gxx gyy gzz] = [1.982 1.979 1.929] (obtained from the simulation of the experimental spectra using Easy spin simulation package[29]). In contrast, the EPR spectra of the hydrogenated-anatase sample show a Ti(III) signature at room temperature. Furthermore, at 90 K the hydrogenated anatase samples show a considerable increase in the Ti(III) signature after illumination[30] – a high intensity Ti(III) signal is observed after 90 minutes illumination with a Xe-lamp (Fig.3a). The simulation indicates an axial-symmetric Ti(III) ion with g values [gzz gxx gyy] = [1.960 1.905 1.905]. Also other EPR features are in line with the observed formation of H2 evolution centers. For instance, these EPR signatures do not occur in rutile, nor are they induced by other reductive treatments (SI, Fig. S11). The fact that XPS does not reveal the formation of surface $Ti^{3+}$ species in substantial amounts (SI, Fig. S10), indicates that they are present only in a subsurface configuration or in concentration below the detection limit of XPS ($\approx$ 1 at%).

Fig. 3b compares the room temperature PL from hydrogenated anatase (Hy-an1) with non-hydrogenated particles (An1). In both cases, a main PL peak around 600-700 nm is visible. This emission is typical for TiO2 nanoparticles[31-33] and is a superposition of trapped-exciton and various defect-related emission bands in the anatase phase. After H2 treatment, the overall intensity of this emission is slightly increased and its maximum is shifted to longer wavelengths. A most significant difference for the H2 treated sample is, however, observed for the PL in the range of 400-450 nm; this peak may be related to self-trapped exciton recombination,[33-35] where excitons may be localized at neighboring $Ti^{3+}$-$O^-$ sites or to band-to-localized defect transitions. Recent theoretical work, however, shows that such neighboring sites are not stable.[36] Therefore this PL in the range just above 400 nm seems more likely associated to a sub-bandgap-defect-state to band transition. Based on the PL wavelength, defects would have to be located energetically close to the main bands, i.e. at $\Delta E \approx$ 0.2-0.5eV, which is significantly closer than the $\Delta E \approx$ 0.8-1.2eV below the conduction band reported for typical $Ti^{3+}$ states[36,37]). In this context, it is interesting that some studies on CVD deposited suboxide layers of TiOx (x<2) find states or even a band of states very close to the anatase conduction band. This finding was supported by DFT calculations and it was reported that these states yield a higher photocatalytic activity for dye decomposition[38].

The above experimental findings therefore clearly point towards a specific defect-configuration introduced by the H2 treatment in anatase. In theory and experiment several sub-band gap bulk and surface electronic states have been reported to mediate the reactivity of TiO2 in classic catalytic as well as in photocatalytic applications, and various defect configurations were reported to show a different stability in anatase and rutile structures.[6,33,36,37] For example, for rutile, H-treatment (under oxygen depleted conditions) is reported to lead preferentially to Ti-interstitials, whereas oxygen vacancy formation is favored in anatase.[37] Created oxygen vacancies are expected to rearrange to take a surface or a subsurface configuration depending on the structure, faceting and the specific treatment. [33,37,39,40] Considering this, the number of configurations of oxygen vacancy / Ti3+ combinations, that can be theoretically stable in anatase within a small energy range, is very high (for example recently for anatase 49 similarly stable Ti3+ defect configurations have been reported[36]). Furthermore, the findings of the present work indicate that the creation of an efficient intrinsic co-catalytic H2 evolution activity on anatase is only achieved for a treatment in pure H2 at elevated temperature. I.e. the common assignment of features from classic reductive treatments to oxygen vacancies, Ti3+ states, Ti-interstitials, and surface reconstruction seems not applicable in a straightforward manner. [2,6,33,36,37]

In summary, the present work demonstrates a remarkable activation of commercial anatase and anatase-rutile powders for photocatalytic H2 evolution after hydrogenating the particles under high pressure, high temperature conditions. The observed experimental findings on the character of the self-induced catalytic centers resemble to a large extent the phenomenology of noble metal co-catalytic effects on TiO2 (such as the polymorph specificity or synergistic effects of anatase and rutile). In spite of some uncertainty about mechanistic details of this activation of anatase powders, EPR, PL and TEM indicate that the high pressure hydrogenation treatment supports the formation of voids in anatase nanoparticles and the formation of a specific RT stable defect structure that appears to be key for the observed co-catalytic effect on TiO2 nanoparticles.

Acknowledgements

The authors would like to acknowledge ERC, DFG and the Erlangen DFG cluster of excellence (EAM) for financial support, and U. Gesenhues (Sachtleben Gmbh) for valuable discussion and providing sample materials. B. W. gratefully acknowledges the Research Training Group "Disperse Systems for Electronic Applications" (GRK 1161).

Keywords: TiO2 • hydrogenation • co-catalyst • water splitting• Ti(III)

**Figure captions:**

**Fig. 1** Photocatalytic hydrogen evolution rate from $TiO_2$ particle suspensions under AM 1.5 (100 mW/cm$^2$) illumination. a) comparison of different anatase, rutile and mixed phase powder before and after hydrogenation (Hy-) at 500 ˚C, 20 bar treatment. b) Hydrogen production rate from hydrogenated anatase samples (Hy-an1) over 5 days of continuous AM 1.5 illumination.

**Fig. 2** a) X-ray diffraction spectra (XRD) of different anatase, rutile and mixed phase samples used in this work; b) Rietveld refinement plot of anatase (An1) and hydrogenated anatase (Hy-an1); c) HRTEM images of anatase (An1) before and after hydrogenation, showing no significant widening of the amorphous shell around particle by treatment. d) TEM bright field images taken under focus (left), underfocus (center) and overfocus (right) conditions for anatase (An1) and hydrogenated anatase (Hy-an1), showing characteristic Fresnel contrast indicating the presence of inner grain voids (arrows).

**Fig. 3** a) EPR spectra for anatase (An1) and hydrogenated anatase (Hy-an1) in dark at room temperature and at 90 K after 90 min illumination. The simulations are in red lines: for An1:g values [1.982 1.979 1.929], g strain [0.020 0.0057 0.0068]; for Hy-an1: g values [1.960 1.905 1.905], g strain [0.070 0.090 0.080]; b) photoluminescence of anatase (An1) and hydrogenated anatase (Hy-an1) in air using 375 nm excitation.

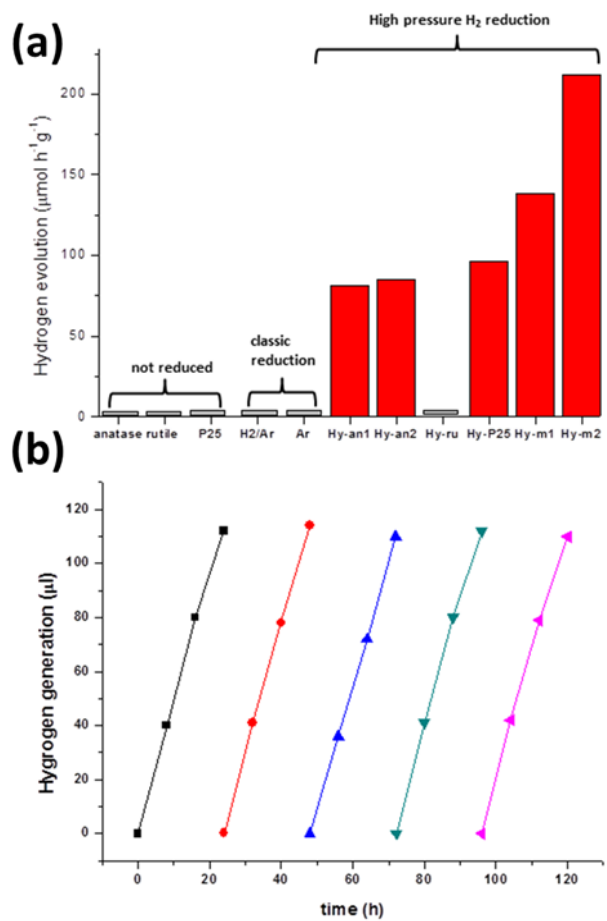

Fig. 1

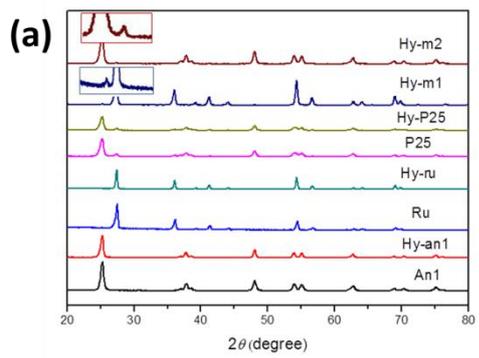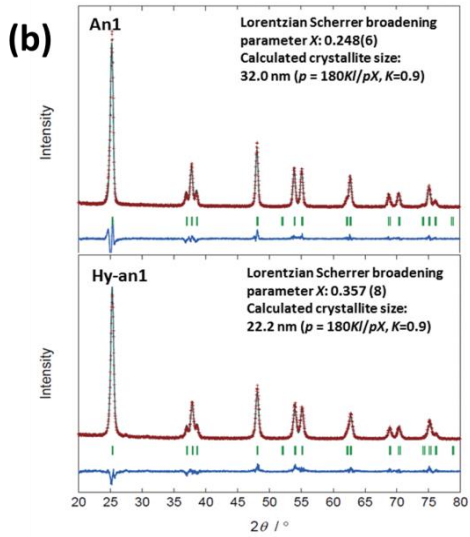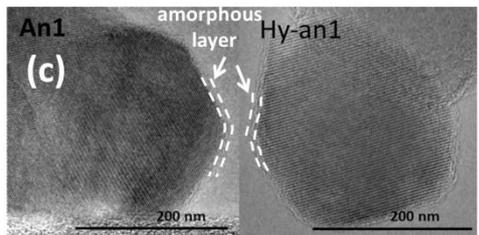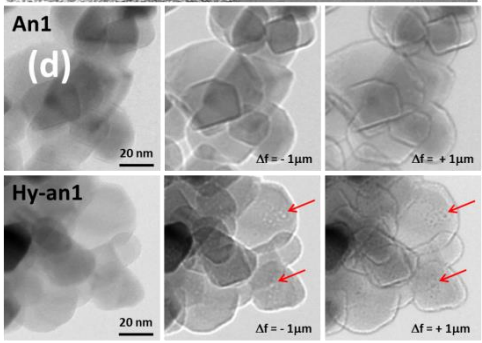

Fig. 2

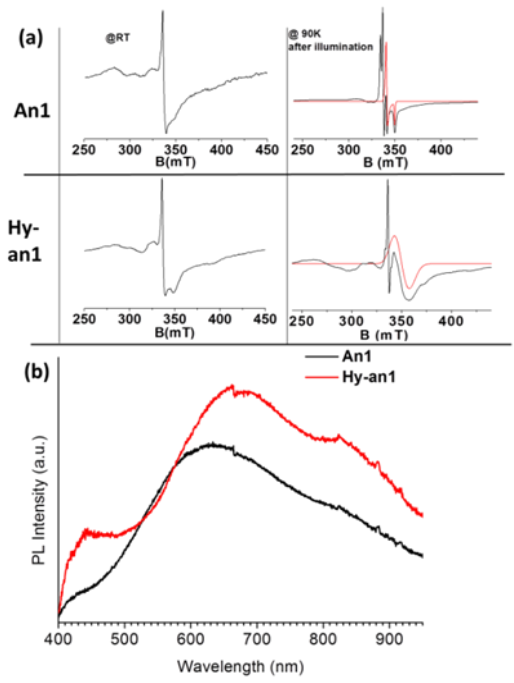

Fig. 3

# Supporting information

# Hydrogenated Anatase: Strong Photocatalytic H$_2$ Evolution Without the Use of a Co-catalyst

Ning Liu[1], Christopher Schneider[1], Detlef Freitag[2], Umamaheswari Venkatesan[3], V. R. Reddy Marthala[3], Martin Hartmann[3], Benjamin Winter[4], Erdmann Spiecker[4], Andres Osvet[5], Eva Zolnhofer[6], Karsten Meyer[6], Tomohiko Nakajima[7], Xuemei Zhou[1], Patrik Schmuki[1]*

[1]Department of Materials Science WW-4, LKO, University of Erlangen-Nuremberg, Martensstrasse 7, 91058 Erlangen, Germany;

[2]High Pressure Laboratory, Chair of Separation Science and Technology, University of Erlangen-Nuremberg, Haberstrasse 11, 91058 Erlangen, Germany;

[3]ECRC - Erlangen Catalysis Resource Center, University of Erlangen-Nuremberg, Egerlandstrasse 3, 91058 Erlangen, Germany;

[4]Center for Nanoanalysis and Electron Microscopy (CENEM), University of Erlangen-Nuremberg, Cauerstrasse 6, 91058 Erlangen, Germany;

[5] Department of Materials Sciences 6, iMEET, University of Erlangen-Nuremberg, , Martensstrasse 7, 91058 Erlangen, Germany;

[6]Inorganic & General Chemistry, University of Erlangen-Nuremberg, Egerlandstrasse 1, 91058 Erlangen, Germany;

[7]National Institute of Advanced Industrial Science and Technology, Tsukuba Central 5, 1-1-1 Higashi, Tsukuba, Ibaraki 305-8565, Japan

*Corresponding author. Tel.: +49 91318517575, fax: +49 9131 852 7582

Email: schmuki@ww.uni-erlangen.de

# Experimental:

As a precursor for hydrogenation, we used $TiO_2$ powders in different crystalline polymorphs (anatase, rutile and mixed phase), different particle sizes (in the range of 20-200 nm) and different purities supplied from different companies (compiled in Table S1 and Fig. S1).

For high pressure $H_2$ treatment, $TiO_2$ powders were annealed in hydrogen atmosphere at 20 bar at 500 °C for 1 h - 3 days. The preparation was performed in a 1 L autoclave (70 mm inner diameter) filled with ultrahigh purity hydrogen gas (purity 99.999 %, Linde). For $Ar/H_2$ or Ar treatments, $TiO_2$ powders were annealed in a tube furnace using $H_2/Ar$ (5%) or Ar flow at 500 °C for 1 h at a flow rate of 6 L/h .

Photocatalytic hydrogen generation was measured under open circuit conditions from an aqueous methanol solution (50 vol%) under AM 1.5 (100 mW/cm$^2$) solar simulator illumination. The amount of $H_2$ produced was measured using a Varian gas chromatograph with a TCD detector. For rate determination, data were taken approximately every 24 h during solar simulator irradiation. To prepare suspensions for $H_2$ measurements, 2 mg $TiO_2$ powders were dispersed in 10 mL of DI water/methanol (50/50 v%) with ultrasonication for 30 min. During illumination, the suspensions were continuously stirred. To test the stability of catalyst for hydrogen evolution, repeated cycle measurements of hydrogen evolution for hydrogenated anatase (Hy-an1) as photocatalyst over a testing period of 5 days (measured in intervals of 8 h) were carried out. The powder remained the same for each cycle.

Regarding the influence of grain size on $H_2$ evolution activity we compared An1 and An2 (particles An1 have a diameter of ≈25 nm while particles An2 have a diameter of ≈200 nm). Both lead, after hydrogenation, to a comparable $H_2$ production rate; i.e., in our experiments we did not find the grain size, within the used variations, to be of a major factor. This may be ascribed to a commonly observed aggregation of $TiO_2$ nanoparticles in neutral aqueous suspensions.

A Hitachi FE-SEM S4800 was used for morphological characterization of the samples. XRD patterns were collected using an X'pert Philips PMD diffractometer with a Panalytical X'celerator detector, using graphite-monochromatized CuKa radiation ($\lambda$ = 1.54056Å). Rietveld refinement was carried out as described in S5.

High-resolution transmission electron microscopy (HRTEM) was performed using an image-side aberration corrected FEI Titan³ 80-300 transmission electron microscope (TEM) operated at an acceleration voltage of 200 kV. Selected area electron diffraction (SAED) patterns were acquired with a Philips CM 300 UT TEM at an acceleration voltage of 300 kV. The powder samples were diluted with deionized water. After 3 minutes of ultrasonication, the dispersions were drop-casted onto TEM copper grids coated with a lacey carbon film and dried overnight before investigation.

The room temperature CW EPR spectra were recorded on an X-band ($v_{mw}$ = 9.84 GHz) EMXmicro BRUKER spectrometer and the 90 K X-band CW EPR spectra were recorded on an ($v_{mw}$ = 9.47 GHz) EMX Plus BRUKER spectrometer using an Oxford flow cryostat with liquid $N_2$ flow. The $B_0$ modulation amplitude used was 0.4 mT, and the modulation frequency was adjusted to $v_{mod}$ = 100 kHz. The microwave power used was low enough to prevent the saturation of the spin systems. Illumination experiments were carried out using a 150W Xe-lamp (Oriel 68806 Basic Power Supply).

The photoluminescence (PL) of the powder samples was excited with a 375 nm diode laser and the spectra were recorded at room temperature with an iHR320 monochromator and Synergy Si CCD camera (both Horiba Jobin-Yvon). The spectra are corrected for the spectral sensitivity of the setup, determined with the help of a calibrated halogen lamp.

[1]H MAS NMR measurements were performed on Agilent DD2 500 MHZ WB spectrometer at a resonance frequency of 499.86 MHz using a 3.2mm MAS NMR probe with a sample spinning rate of 15kHz. In order to suppress the [1]H background signals arising from the probe head and rotor caps, spectra were recorded using a DEPTH one pulse sequence. In this sequence, a 90° pulse length of 3.1 µs followed by two 180° pulse lengths of 6.2 µs and a recycle delay of 2s were applied. A total of 200 scans were accumulated for each spectrum.

**Table S1** Overview of TiO$_2$ powders used in this work.

| Name | Company | Nominal composition | Remarks |
|---|---|---|---|
| An1 | Aldrich | anatase | Purity: 99.8%, particle size: 25-35 nm |
| An2 | Sachtleben | anatase | Purity: 99.8%, particle size: 100-200 nm |
| An3 | Aldrich | anatase | Purity: 99.9%, particle size: 100-200 nm |
| Ru | Sachtleben | rutile | Purity: 99.8%, particle size: 200-300 nm |
| P25 | Degussa | mixed | Purity: 99.8%, particle size: 20 nm |
| M1 | Aldrich | mixed | Purity: 99.8%, particle size: 25-35 nm |
| M2 | Aldrich | mixed | Purity: 99.8%, particle size: 25-35 nm |
| M3 | Sachtleben | mixed | Purity: 99.8%, particle size: 200-400 nm |
| Hy-an1 | Aldrich | anatase | Hydrogenated at 500 ˚C, 20 bar for 1h |
| Hy-an2 | Sachtleben | anatase | Hydrogenated at 500 ˚C, 20 bar for 1h |
| Hy-ru | Sachtleben | rutile | Hydrogenated at 500 ˚C, 20 bar for 1h |
| Hy-m1 | Aldrich | mixed | Hydrogenated at 500 ˚C, 20 bar for 1h |
| Hy-m2 | Aldrich | mixed | Hydrogenated at 500 ˚C, 20 bar for 3 days |
| Hy-P25 | Degussa | mixed | Hydrogenated at 500 ˚C, 20 bar for 1h |
| HA-an1 | Aldrich | anatase | Ar/H$_2$ stream at 500 ˚C for 1 h |
| Ar-an1 | Aldrich | anatase | Ar stream at 500 ˚C for 1 h |

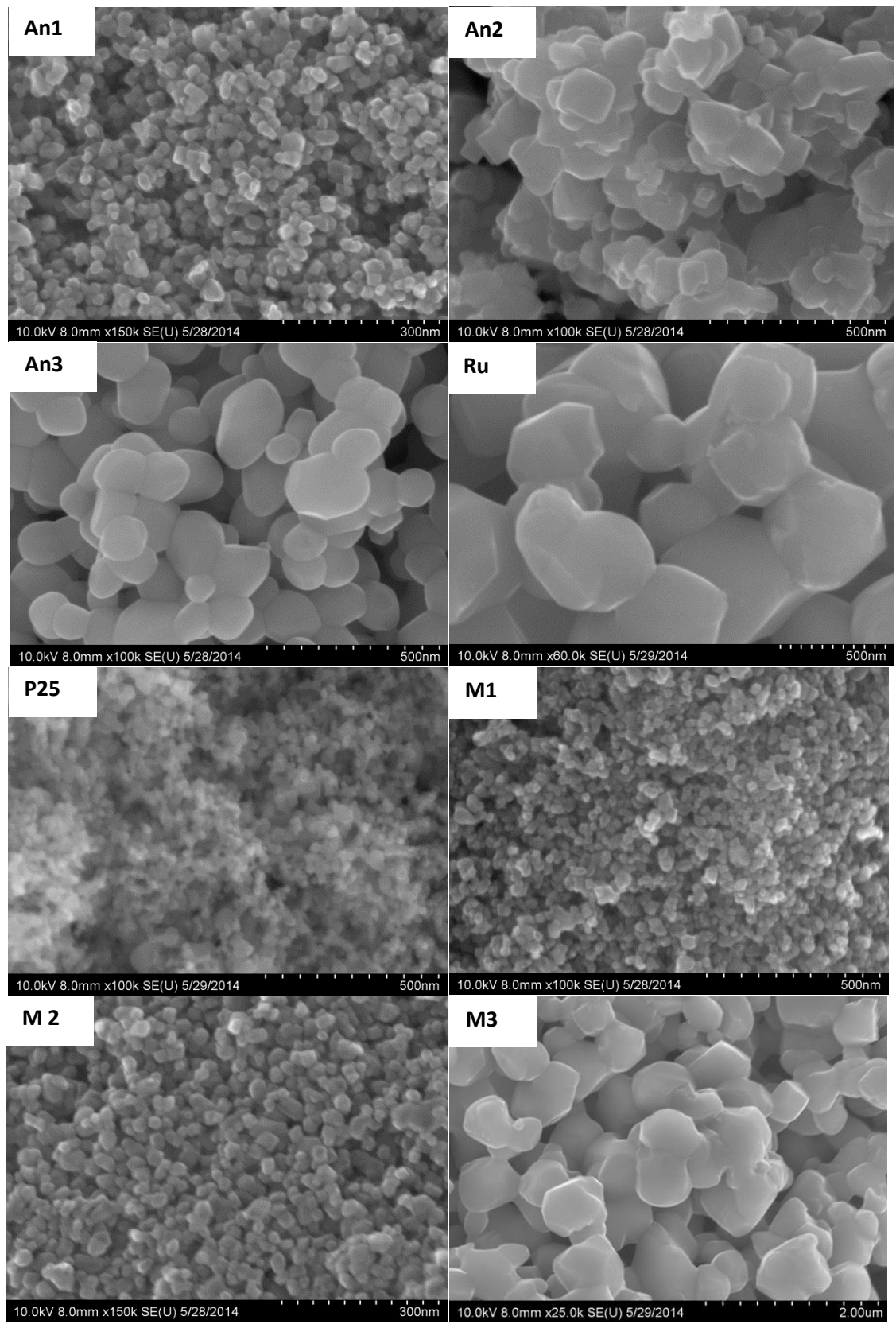

Fig. S1 SEM images of different TiO$_2$ powders used in this work (diameters were evaluated from averaging measurements on > 10 grains).

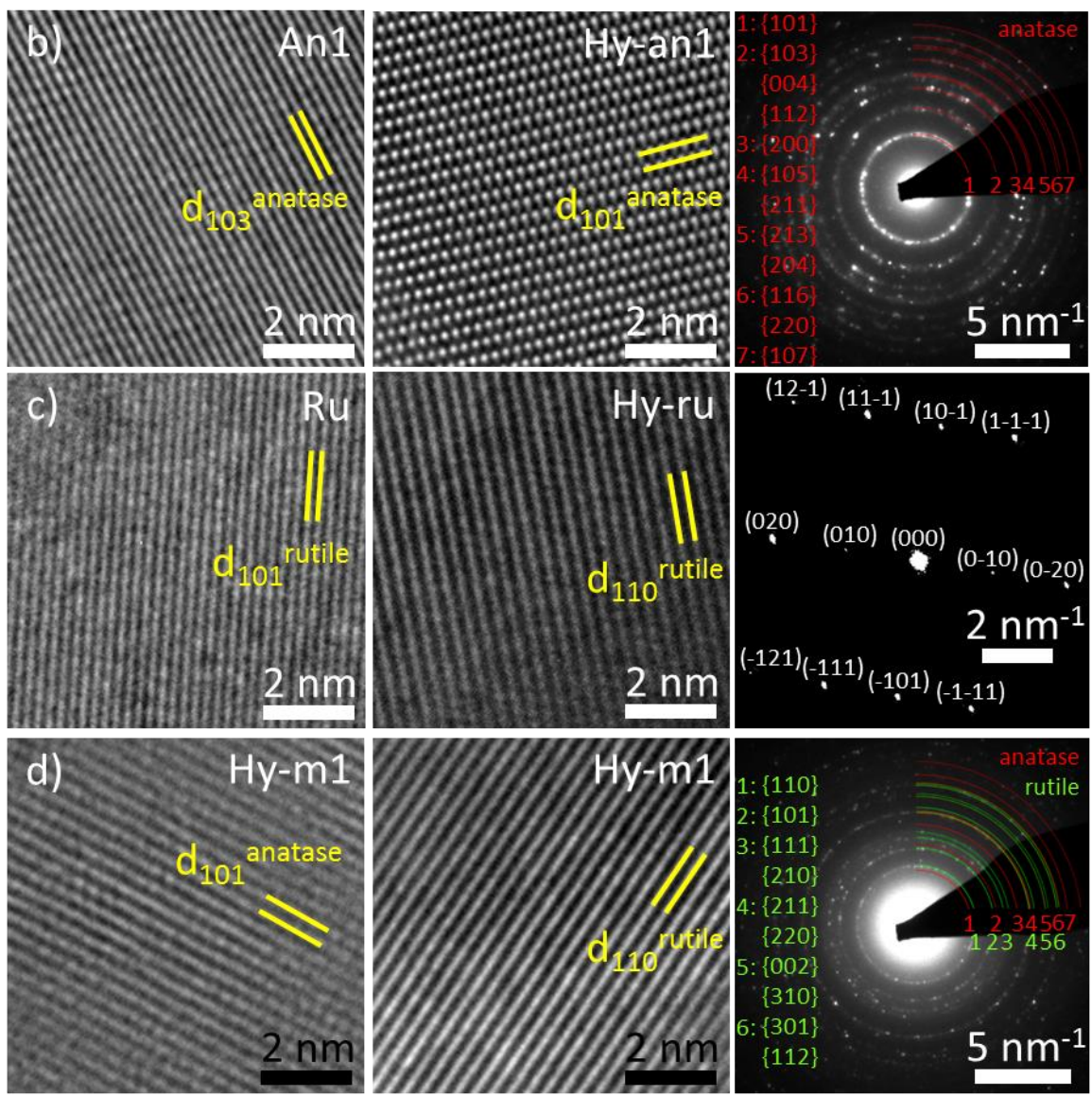

Fig. S2 High resolution (HRTEM) images of selected anatase, rutile and mixed phase samples before and after hydrogenation (left) and selected area diffraction pattern (right) for corresponding hydrogenated samples. In the SAED patterns of the hydrogenated samples expected diffraction rings for the anatase phase of (b) are indicated in red; diffraction pattern in (c) are taken with particle oriented in zone axis [101]. The HRTEM images of the hydrogenated mixed sample (d) shows one particle with anatase and one with rutile crystal structure. The SAED ring pattern exhibits both expected phases with a higher amount of rutile particles (expected ring pattern of anatase in red and rutile in green; legend of anatase rings see (b)). Respective lattice spacings are indicated with yellow lines: $d_{101}^{anatase}$ = 0.352 nm, $d_{103}^{anatase}$ = 0.243 nm, $d_{110}^{rutile}$ = 0.325 nm, $d_{101}^{rutile}$ = 0.249 nm.

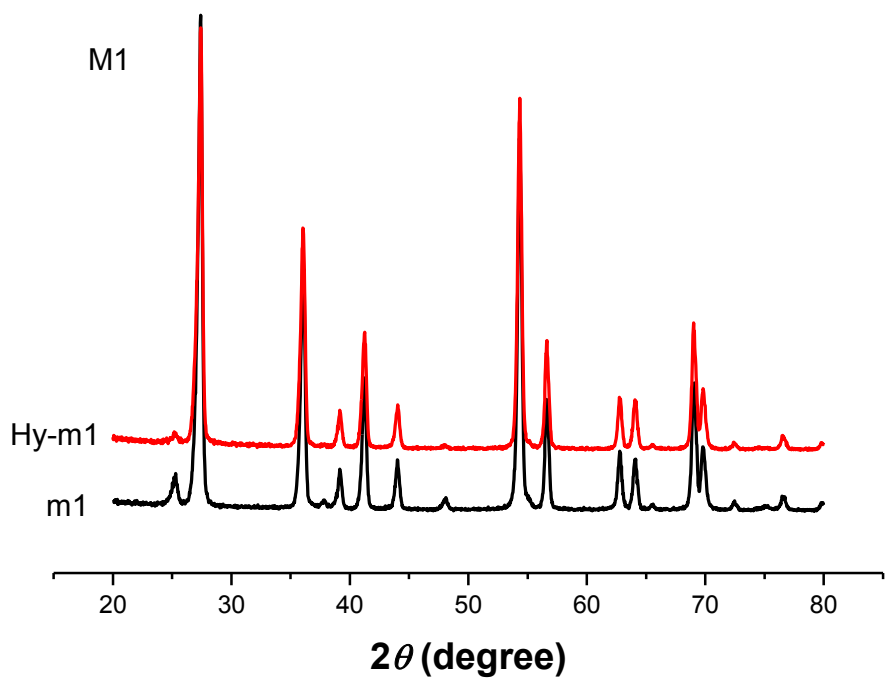

Fig. S3 XRD patterns for mixed phase particles before (black) and after hydrogenation (red), that show a decrease of the anatase content from approx. 12% to 4% after the $H_2$ treatment.

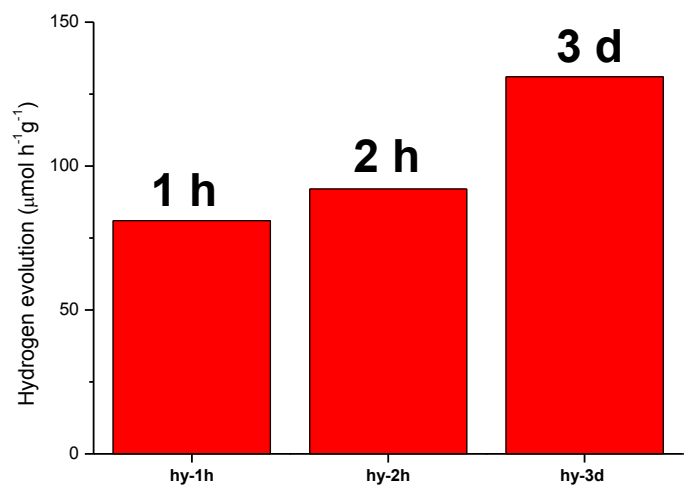

Fig. S4 Photocatalytic hydrogen evolution rates from An1 suspensions under AM 1.5 (100 mW/cm$^2$) illumination for different hydrogenation durations (1 h, 2 h, and 3 days).

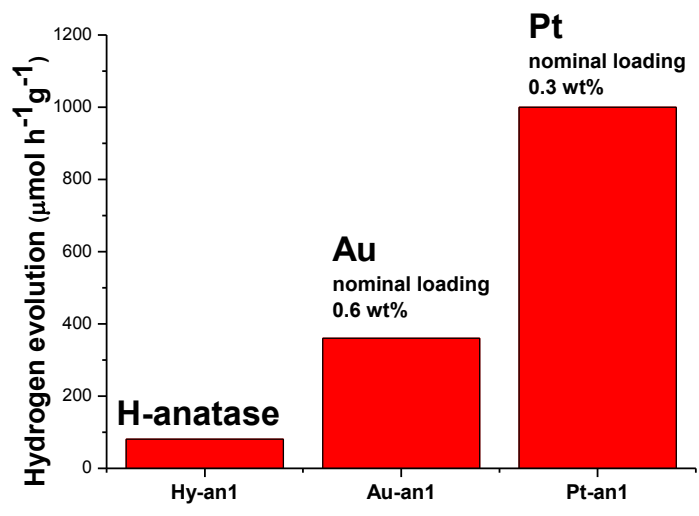

Fig. S5 Comparison of the hydrogen evolution efficiency for the hydrogenated anatase particles H-anatase (Hy-an1) with same type nanoparticle (an1) loaded with noble metal co-catalyst (Pt and Au).

Rietveld analysis of XRD spectra was carried out using the RIETAN-FP program and Thompson-Cox-Hastings (TCH) profile function to calculate the crystallite size. The hydrogenated $TiO_2$ and $TiO_2$: H were fitted by using a $TiO_2$ anatase model. Atom displacement parameter B for Ti and O atoms of $TiO_2$:H were fixed to be the refined parameter. Very slight lattice parameter variation and clear difference of nominal crystallite size (as given below) were observed.

**$TiO_2$**

Structural model: Anatase $TiO_2$
Space group: $I\,4_1/a\,m\,d$ (VOL. A, 141)
$R_{WP}$ = 13.812%, $R_e$ = 7.671%
$a$ = 3.7835(1) Å, $c$ = 9.5110 (4) Å, $V$ = 136.146 (9) Å$^3$

| Atom | $x$ | $y$ | $z$ | $B$ |
|---|---|---|---|---|
| Ti | 0 | 0 | 0 | 0.40(6) |
| O | 0 | 0 | 0.2093(3) | 0.71(9) |

Profile function: TCH's pseudo-Voigt function
Lorentzian Scherrer broadening parameter $X$: 0.248(6)
Calculated crystallite size: 32.0 nm ($p = 180K\lambda/\pi X$, $K$=0.9)

**Hydrogenated $TiO_2$**

Structural model: Anatase $TiO_2$
Space group: $I\,4_1/a\,m\,d$ (VOL. A, 141)
$R_{WP}$ = 13.206%, $R_e$ = 9.252%
$a$ = 3.7845(1) Å, $c$ = 9.5017(5) Å, $V$ = 136.086 (12) Å$^3$

| Atom | $x$ | $y$ | $z$ | $B$ |
|---|---|---|---|---|
| Ti | 0 | 0 | 0 | 0.40 |
| O | 0 | 0 | 0.2094(3) | 0.71 |

Profile function: TCH's pseudo-Voigt function
Lorentzian Scherrer broadening parameter $X$: 0.357 (8)
Calculated crystallite size: 22.2 nm ($p = 180K\lambda/\pi X$, $K$=0.9)

Fig. S6 Rietveld refinement of anatase powder An1 and hydrogenated anatase Hy-an1.

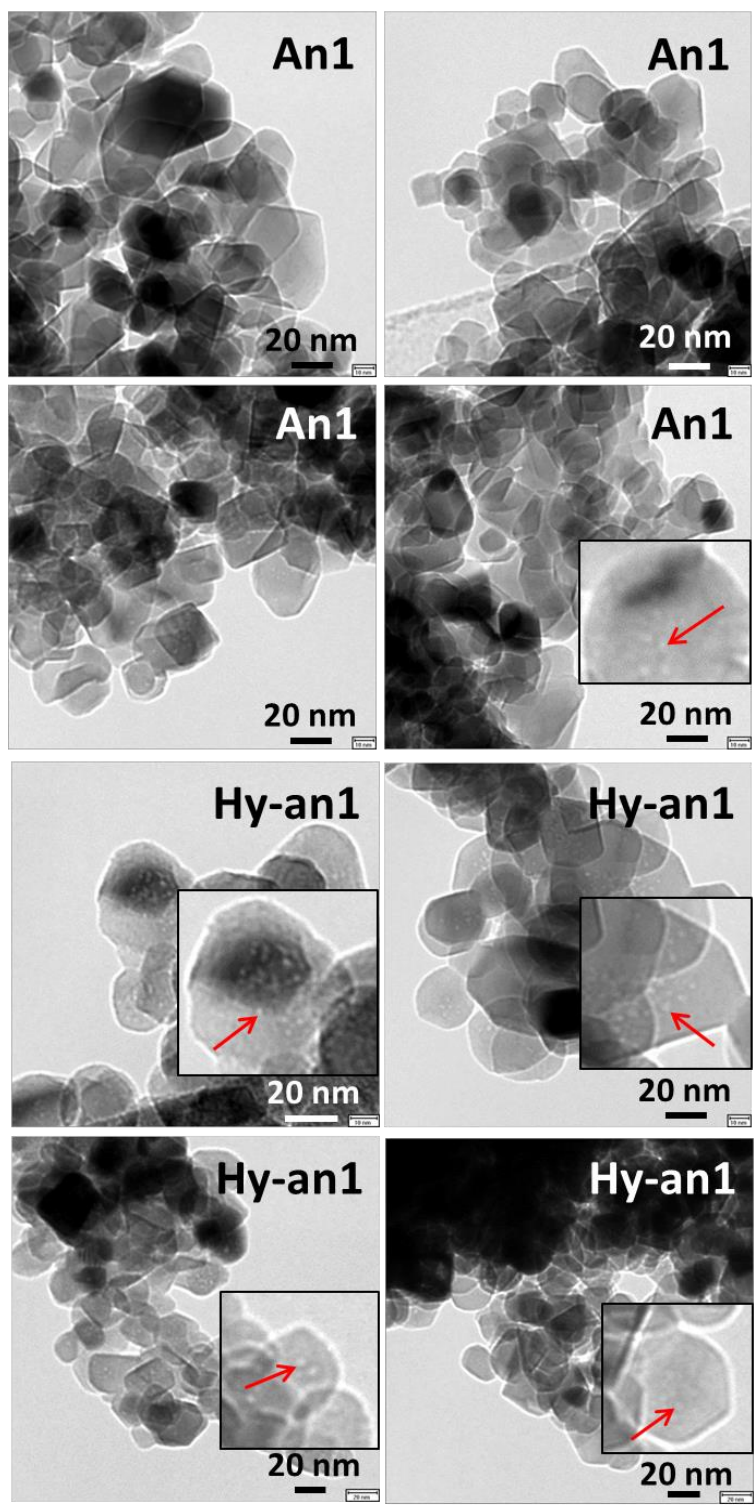

Figure S7 Additional TEM bright field images taken under Fresnel contrast conditions (underfocus values between -1 μm and -2 μm for anatase (An1) and hydrogenated anatase

(Hy-an1) particles. Clear voids occur strongly in Hy-an1 whereas they only very rarely can be found in An1 (see arrows). (Note that slightly different magnifications have been used.)

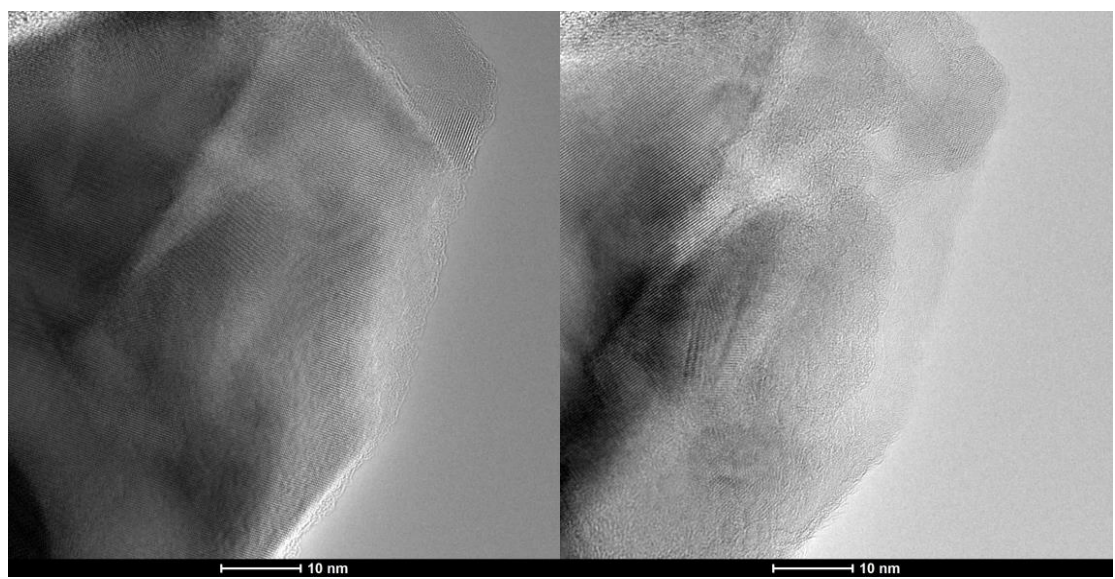

Figure S8. Exclusion of possible artifacts by HRTEM: Sequence of HRTEM images illustrating that long time exposure (right) rather leads to amorphization of anatase NP than to crystallization. I.e. HRTEM induced crystallization in the present investigation can be excluded as an artifact (e.g. affecting [reducing] the amorphous shells around particles).

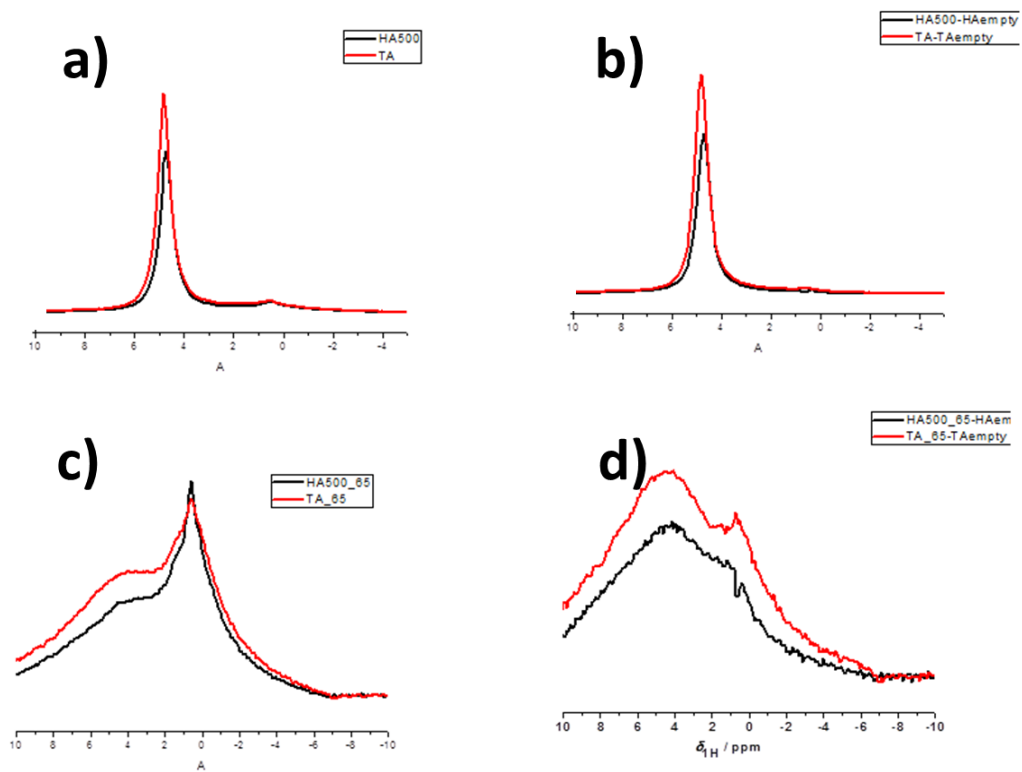

Fig. S9 $^1$H magic angle NMR spectra of anatase particles with (black) and without (red) high pressure $H_2$ treatments. Before (a, b) and after (c, d) 65 °C heating treatment to reduce influence of surface adsorbed water (pick-up during sample transfer from hydrogenation set-up to spectrometer). Spectra a) and c) are taken before subtraction of background signals (rotor caps, probe head). Spectra b) and d) after subtraction of these artifacts - illustrating that sharp peaks around δ=1ppm must be ascribed mainly to machine- specific artifacts (in line with [S1]).

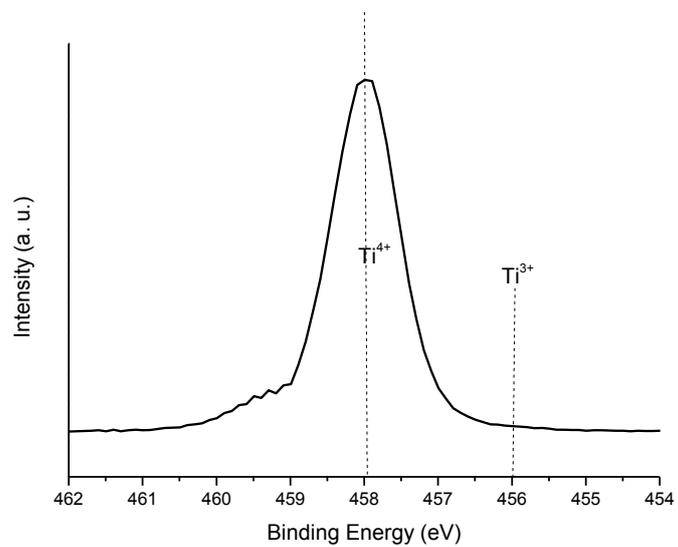

Fig. S10 XPS spectra of Ti2p for hydrogenated anatase, indicating that with accuracy of XPS ($\approx$ 1 at%) surface $Ti^{3+}$ cannot be detected.

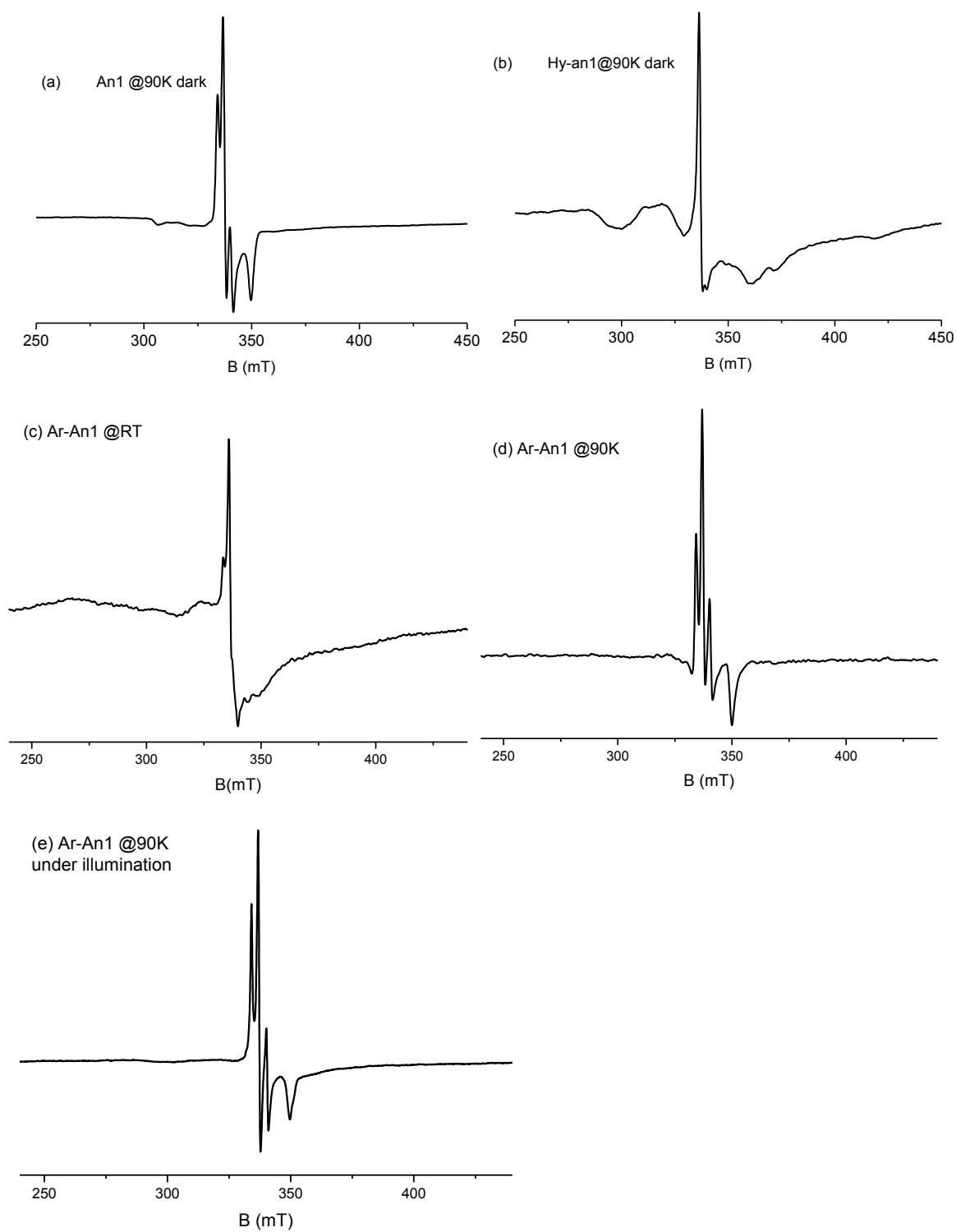

Fig. S11 Further EPR reference measurements for anatase (An1) (a) and hydrogenated anatase (Hy-an1) (b) in dark at 90 K and Ar treated An1 at room temperature (c), at 90 K (d), after 90 min illumination (e).

Supplementary references: